\documentclass[12pt]{article}
\usepackage{dcolumn}
\newcounter{saveeqn}%
\newcommand{\alpheqn}{\setcounter{saveeqn}{\value{equation}}%
\stepcounter{saveeqn}\setcounter{equation}{0}%
\renewcommand{\theequation}
{\mbox{\arabic{saveeqn}-\alph{equation}}}}%
\newcommand{\reseteqn}{\setcounter{equation}{\value{saveeqn}}%
\renewcommand{\theequation}{\arabic{equation}}}%
\begin{document}
\vspace*{1.3in}
\begin{center}
{\large{\bf Net-baryon scaling near the QCD critical point\\}}
\vspace{0.4in}
N.G. Antoniou\footnote{e-mail:nantonio@cc.uoa.gr}, F.K. Diakonos and
A.S. Kapoyannis\\
\smallskip
{\it Department of Physics, University of Athens,\\
GR-15771 Athens, Greece\\}
\date{May 2001}
\end{center}

\vspace{0.4in}
\begin{abstract}
The net-baryon density at midrapidity is proposed as an order parameter
in the search for the QCD critical point in heavy ion collisions.
As a function of the initial energy and the total number of participants,
this quantity obeys a scaling law, dictated by the critical exponents
of the appropriate universality class. The corresponding scaling variable
specifies the proximity of a given experiment to the critical point. Within
this framework, measurements at the SPS are discussed and predictions for
RHIC and LHC are extracted.
\end{abstract}

\newpage
\setcounter{page}{1}

\section{Introduction}

A remarkable property of the QCD phase diagram $(\rho~-~T)$ is the existence
of an endpoint along the critical line of the first order quark-hadron
phase transition \cite{Wilc}. It defines a critical point of second order,
belonging to the universality class of a $3d$ Ising system and located on a
line of nonzero baryonic density $\rho=\rho_c$.
The QCD critical point is associated with the chiral phase transition
in the sense that it is the remnant of a tricritical point corresponding
to the chiral limit $m_u=m_d=0$ \cite{Step1}.
In other words, the existence of a second-order critical point, at
nonzero baryonic density, is a fundamental property of real QCD with
small but nonzero quark masses ($m_u,m_d \neq 0$).                                      
The QCD critical point communicates with the hadronic world through
the fluctuations of a scalar field ($\sigma$-field)
which carries the quantum numbers of an isoscalar ($\sigma$-meson)
as the manifestation of a quark condensate, $\sigma \sim \langle \bar{q}q
\rangle$, in thermal environment. At the critical temperature (and
infinite volume)
the isoscalar has zero mass, in order to provide the infinite wavelength
mode required by the divergence of the correlation length. It remains,
therefore, stable in the thermal environment of a nuclear collision as
long as the temperature stays close to the critical value. At the
freeze-out stage the mass of the $\sigma$-field 
may reach the two-pion threshold $(m_{\sigma} \stackrel{>}{\sim}
2 m_{\pi})$ and become accessible to observation \cite{Step1,Anto1}.
In the effective theory of the QCD critical point, the classical
$\sigma$-field is a natural order parameter, the fluctuations of which obey
scaling laws dictated by the critical exponents of the $3d$ Ising system
($\eta \approx 0, \beta \approx \frac{1}{3}, \delta \approx 5, \nu \approx
\frac{2}{3}$).
In a baryonic environment, however, the chiral condensate is expected to
have, at $T=T_c$, a strong dependence on the net-baryon density, driving the
$\sigma$-field close to zero for $\rho \approx \rho_c$:
$\langle \bar{q} q \rangle_{\rho} \approx \lambda \left(
\frac{\rho - \rho_c}{\rho_c}\right) \langle \bar{q} q \rangle_{0} +
O[(\rho-\rho_c)^2]$
where $\lambda$ is a dimensionless constant of the order of unity
\cite{Broc}. This dependence suggests a new order parameter,
$m=\rho-\rho_c$, associated with the critical properties of the baryonic
fluid created
in a quark-hadron phase transition. In fact, approaching the critical
point in the phase diagram, both the $\sigma$-field
fluctuations and the fluctuations of the order parameter $m(\vec{x})$
obey the same scaling laws ($\langle \bar{q} q \rangle_{\rho} \sim
m(\vec{x})$).

In this work we exploit the scaling properties of the order parameter
$m(\vec{x})$, properly adjusted to measurable quantities,
in heavy ion collisions. For this purpose we consider net-baryon production
in collisions of heavy nuclei with total number of participants $A_t$
and initial energy corresponding to a total size in rapidity $\Delta y = L$.
The paper is organized as follows: In Section 2 we discuss in detail the
scaling properties of the net-baryon fluid near the critical point. In
Section 3 we present shortly phenomenological consequences of the presence
of critical fluctuations in the baryonic sector. Finally in Section 4 we
present our conclusions concerning the possibility to approach and
observe the QCD critical point in heavy-ion colliders.

\section{Net-baryon scaling}

The created baryons in the process of quark-hadron phase transition occupy
a cylindrical volume with transverse radius $R_{\perp} \sim A_{\perp}^{1/3}$
and longitudinal size $L$ (in rapidity). The parameter $A_{\perp}$
specifies the effective number of participants, contributing to the
transverse geometry of the collision, and it is assumed $A_{\perp} \approx
\frac{A_t}{2}$, valid both for central ($A_{\perp}=A_{min}$)
and non-central collisions. Projecting out the net-baryon system onto the
longitudinal direction we end up with a $1d$ liquid confined in a finite
rapidity region of size $L$ with local density $\rho(y)=\frac{n_b(y)}
{\pi R_{\perp}^2 \tau_f}$,
directly related to the measurable net-baryon density in rapidity
$n_b(y)=\frac{dN_b}{dy}$. Putting $R_{\perp}=R_o A_{\perp}^{1/3}$, we
introduce a characteristic volume $V_o=\pi R_o^2 \tau_f$
in terms of the freeze-out time scale $\tau_f \geq 6-8~fm$ \cite{Mull}
and the nuclear-size scale $R_o$ which contains also any growth
effects near the critical point ($R_o \geq 1.2~fm$).                                     ).
Using $V_o^{-1}$ as a scale for baryonic, freeze-out densities, the order
parameter $m(y)$ of the $1d$ baryonic liquid is written:
\begin{equation}
m(y)=A_{\perp}^{-2/3} n_b(y) - \rho_c~~~~~,~~~~~0 \leq y \leq L
\label{eq:eq1}
\end{equation}
In what follows, our basic assumption is that the deconfined phase of quark
matter, in heavy-ion collisions, approaches the QCD critical point in local
thermal equilibrium. In the framework of inside-outside mechanism, this
process is implemented by considering the isothermal space-time hyperbolas
$t^2 - x^2_{\parallel}=\tau^2$ and employing the corresponding rapidity
variable as the appropriate longitudinal coordinate. In this description,
the assumed local equlibrium in the conventional geometry $(x_{\parallel},
{\vec{x}}_{\perp},t)$ is translated as global equilibrium in the new
geometry $(y,{\vec{x}}_{\perp},\tau)$ and one may consistently impose,
near criticality,
static scaling laws on the order parameter $m(y)$. In particular, the
expected singularity at $T=T_c$ is incorporated in a general expansion of the
form:
\begin{equation}
m(y)=t^{\beta} \left[ F_o(y/L) + t F_1(y/L) + ... \right]
\label{eq:eq2}
\end{equation}
where $t \equiv \frac{T_c - T}{T_c}$ and $\beta$ is the
appropriate critical exponent ($\beta \approx \frac{1}{3}$). The leading
term $F_o(y/L)$ in the expansion (\ref{eq:eq2}) has a universal power-law
behaviour near the walls $(y=0,L)$, shared by all fluids belonging to the
same universality class and confined in a finite, one-dimensional region
\cite{Maci,Fish}. At midrapidity $(y \approx \frac{L}{2})$,
eq.(\ref{eq:eq2}) gives the deviation of the measurable, bulk density of net
baryons, from the critical value $\rho_c$, as we approach the critical point,
not along the critical isochore $(\rho = \rho_c)$ but along the freeze-out
line:
\begin{equation}
A_{\perp}^{-2/3} n_b = \rho_c + t_f^{\beta} \left[ F_o + t_f F_1 +
...\right]
\label{eq:eq3}
\end{equation}
where $t_f=\frac{T_c - T_f}{T_c}$, $n_b=n_b(L/2)$, $F_i=F_i(1/2)$ and $T_f$
is the freeze-out temperature.
Integrating now eq.(\ref{eq:eq2}) in the
interval $0 \leq y \leq L$ we obtain at $T=T_f$:
\begin{equation}
A_{\perp}^{-2/3} A_t L^{-1} = \rho_c + t_f^{\beta} (I_o + t_f I_1 + ...)
\label{eq:eq4}
\end{equation}
where $I_i=\int_0^1 F_i(\xi) d\xi$. Introducing the variable
$z_c=A_{\perp}^{-2/3} A_t L^{-1}$ we find from eqs.(\ref{eq:eq3}) and
(\ref{eq:eq4}) a scaling law for the net-baryon density at midrapidity:
\begin{equation}
A_{\perp}^{-2/3} n_b = \Psi(z_c,\rho_c)~~~~~;~~~~~z_c \geq \rho_c
\label{eq:eq5}
\end{equation}
where the scaling function $\Psi(z_c,\rho_c)$ has the property
$\Psi(z_c=\rho_c,\rho_c)=\rho_c$. In the crossover regime $z_c < \rho_c$,
where critical fluctuations disappear, the local density at midrapidity is,
to a good approximation, $n_b \approx A_t L^{-1}$ suggesting a
continuous extension of the scaling law (\ref{eq:eq5}) in this region
with $\Psi(z_c,\rho_c)=z_c$ ($z_c < \rho_c$). It is of interest to note that
although the scaling function $\Psi(z_c,\rho_c)$ is continuous at the
critical point $(z_c=\rho_c)$, the first derivative is expected to be
discontinuous at this point in accordance with the nature of the phase
transition (critical point of second order).

In order to study in detail the critical behaviour of the baryonic fluid
in terms of the new variables $n_b$, $z_c$ a further investigation of
the structure of the scaling function $\Psi(z_c,\rho_c)$ for $z_c \geq
\rho_c$ is necessary. For this purpose we approximate the nearby part of the
freeze-out line, close to the critical point $(t_f \ll 1)$, by truncating
the series (\ref{eq:eq2}) keeping only the next to the leading term
$F_1(y/L)$. As a result, the equations (\ref{eq:eq3}) and (\ref{eq:eq4})
are written correspondingly:
\alpheqn
\begin{eqnarray}
A_{\perp}^{-2/3} n_b &\approx& \rho_c + t_f^{\beta} (F_o + t_f F_1 )
\\
\label{eq:eq6a}
z_c-\rho_c &\approx& t_f^{\beta} (I_o + t_f I_1)
\label{eq:eq6b}
\end{eqnarray}
\reseteqn
Neglecting terms of order $O(t_f^2)$
in the power expansion of the quantity $(I_o + t_f I_1)^{1/\beta}$ and
combining eqs.(6) we finally obtain:
\alpheqn
\begin{eqnarray}
A_{\perp}^{-2/3} n_b &=& \rho_c +
\frac{F_o}{F_1} [f(z_c,\rho_c)]^{\beta} + C [f(z_c,\rho_c)]^{\beta+1}  \\
\label{eq:eq7a}
A_{\perp}^{-2/3} n_b &=& \rho_c + F_o t_f^{\beta} (1 +
t_f \frac{C I_o^{1+\frac{1}{\beta}}}{F_o}) \\
\label{eq:eq7b}
f(z_c,\rho_c) &=& \frac{1}{G}(-1 + \sqrt{1 +
2 G (z_c -\rho_c)^{1/\beta}})~~~;~~~z_c \geq \rho_c
\label{eq:eq7c}
\end{eqnarray}
\reseteqn
where $C=\frac{F_1}{I_o^{1+\frac{1}{\beta}}}$, $G=\frac{2 I_1}{\beta
I_o^{1+\frac{1}{\beta}}}$.

The component $F_o(y/L)$ which dominates the order parameter
$m(y)$ in the limit $T \to T_c$ is approximately constant in the central
region $(y \approx \frac{L}{2}~,~L \gg 1)$, far from the walls (at the points
$y=0,L$) due to the approximate translational invariance of the finite system
in this region. On the other hand, approaching the walls, $F_o(y/L)$
describes the density correlation with the endpoints and obeys a universal
power law. In summary:
\alpheqn
\begin{eqnarray}
F_o(\xi) &\sim& const.~~~~~(\xi \approx \frac{1}{2})  \\
\label{eq:eq8a}
F_o(\xi) &\sim& \xi^{-\beta/\nu}~~~~~(\xi \geq 0)  \\
\label{eq:eq8b}
F_o(\xi) &\sim& (1-\xi)^{-\beta/\nu}~~~~~(\xi \leq 1)
\label{eq:eq8c}
\end{eqnarray}
\reseteqn
The solution which fulfils these requirements is:
\alpheqn
\begin{eqnarray}
F_o(\xi)&=&g[\xi (1- \xi)]^{-\beta/\nu}~~(0 \leq \xi \leq 1) \\
\label{eq:eq9a}
\frac{F_o}{I_o}&=&\frac{4^{\beta/\nu}}{B(1-\frac{\beta}{\nu},
1-\frac{\beta}{\nu})}~~~;~~~B(n,m)=\frac{\Gamma(m)\Gamma(n)}
{\Gamma(m+n)}
\label{eq:eq9b}
\end{eqnarray}
\reseteqn
Inserting the constant $\frac{F_o}{I_o}$ into the 
eq.(\ref{eq:eq7a}), the final form of the scaling function $\Psi(z_c,\rho_c)$
is obtained:
\alpheqn
\begin{eqnarray}
\Psi(z_c,\rho_c)&=&\rho_c+\frac{4^{\beta/\nu}}{B(1-\frac{\beta}{\nu},
1-\frac{\beta}{\nu})} \left[f(z_c,\rho_c)\right]^{\beta} \nonumber\\
&& + C \left[f(z_c,\rho_c)\right]^{\beta +1}~~~~;~~~~z_c \geq \rho_c \\
\label{eq:eq10a}
\Psi(z_c,\rho_c)&=& z_c~~;~~z_c < \rho_c
\label{eq:eq10b}
\end{eqnarray}
\reseteqn
where $f(z_c,\rho_c)$ is given by eq.(\ref{eq:eq7c}). It is straightforward to
show that the discontinuity of the first derivative of $\Psi(z_c,\rho_c)$ at
$z_c=\rho_c$ is a nonzero universal constant:
$disc\left(\frac{d \Psi}{d z_c}\right)_{\rho_c} \approx 1 - \frac{2}{\pi}$
($\frac{\beta}{\nu} \approx \frac{1}{2}$) as expected from the
characteristic properties of a second-order phase transition.
Combining eqs.(\ref{eq:eq5}-10) one may propose a framework for
the treatment of certain phenomenological aspects of the QCD critical point.
The scaling law (\ref{eq:eq5}) combined with eq.(10) involves
three nonuniversal parameters: (a) the critical density $\rho_c$ and (b) the
constants $C,G$ which give a measure of the nonleading effects, allowing
to accomodate in the scaling function (10) processes not very
close to the critical point. We have used measurements at the SPS in order
to fix these parameters on the basis of eqs.(\ref{eq:eq5}) and
(10). More specifically, in a series of experiments (Pb+Pb, S+Au,
S+Ag, S+S) with central and noncentral (Pb+Pb) collisions at the SPS
\cite{NA35,Coop} net baryons have been measured at midrapidity whereas the
scaling variable $z_c = A_{\perp}^{-2/3} A_t L^{-1}$, associated with these
experiments, covers a sufficiently wide range of values $(1 \leq z_c \leq 2)$
allowing for a best fit solution. The outcome of the fit is consistent with
the choice $G \approx 0$ and the equations (7) are simplified as follows:
\alpheqn
\begin{eqnarray}
A_{\perp}^{-2/3} n_b &=& \rho_c + \frac{2}{\pi} (z_c-\rho_c)
+ C (z_c-\rho_c)^4 \\
\label{eq:eq11a}
A_{\perp}^{-2/3} n_b &=& \rho_c + \frac{2 I_o}{\pi} t_f^{1/3} (1 +
t_f \frac{\pi C I_o^{1/3}}{2}) \\
\label{eq:eq11b}
A_{\perp}^{-2/3} n_b(y) &=& \rho_c +
\frac{(z_c-\rho_c)L}{\pi \sqrt{y(L-y)}} + O[(z_c - \rho_c)^4]
\label{eq:eq11c}
\end{eqnarray}
\reseteqn
In eqs.(11) we have used the approximate values of the critical
exponents $\beta \approx \frac{1}{3}$, $\frac{\beta}{\nu} \approx
\frac{1}{2}$ in the $3d$ Ising universality class \cite{Ma}. We have also
added eq.(\ref{eq:eq11c}) which gives the universal behaviour of the
net-baryon density $n_b(y)$, in the vicinity of the critical point
$(z_c -\rho_c \ll 1)$. The fitted values of the parameters in
eq.(\ref{eq:eq11a}) are $\rho_c =0.81$, $C=0.68$
and the overall behaviour of the solution is shown in
Fig.~1. In turns out that the critical density is rather small, compared
to the normal nuclear density $\rho_o \approx 0.17~fm^{-3}$: $\rho_c \leq
\frac{\rho_o}{5}$ ($R_o \geq 1.2~fm$, $\tau_f \geq 6~fm$),
suggesting that the critical temperature remains close to the value
$T_c \approx 140~MeV$
obtained in studies of QCD on the lattice at zero chemical potential
\cite{Laer}.
The difference $d_c=\vert z_c -\rho_c \vert$ 
in Fig.~1 is a measure of the proximity of a given experiment $(A_t,L)$
to the critical point. We observe that the central Pb+Pb collisions at
the SPS ($d_c \approx 1.1$) drive the system into the most distant freeze-out
area, from the critical point, as compared to other processes at the same
energy. In fact, the most suitable experiments to bring quark matter close
to the critical point at the SPS are: S+S $(d_c \approx 0.18)$, S$_i$+S$_i$
$(d_c \approx 0.20)$ and C+C $(d_c \approx 0.06)$                          ),
central collisions. One may even reach the critical point $(d_c \approx 0)$
at the SPS with medium-size nuclei, either in noncentral collisions:
S+S, S$_i$+S$_i$ ($A_t \approx 29$) or at lower energies:
C+C $(P\approx 130~A \frac{GeV}{c})$. Also, the experiments at RHIC, with
central Au+Au $(d_c \approx 0.27)$ collisions, come close to the critical
point and may even reach it exactly ($d_c \approx 0$) either with lighter
nuclei or with noncentral Au+Au collisions ($A_t \approx 165$).
Finally at the LHC $(\sqrt{s} \approx 5.5~TeV)$,
Pb+Pb collisions are expected to drive quark matter into the crossover
area ($z_c < \rho_c$, $d_c \approx 0.13$) where no critical fluctuations
occur.
At lower energies however ($\sqrt{s} \approx 1.4~TeV$), Pb+Pb central
collisions may
reach the critical point even at the LHC. Obviously, in order to have
a clear answer on the proximity of a given experiment to the critical point,
precision measurements of the net-baryon density at midrapidity, in a wide
range of energies (between SPS and RHIC), are needed, both for central and
noncentral collisions. With such measurements, the scaling law (\ref{eq:eq5})
may be
tested in its full extent and a sharp determination of the critical density
can be achieved through eq.(7a).

The freeze-out line ($n_b$ versus $T_f$) given by eq.(11b) is
shown in Fig.~2. We have used the freeze-out temperatures for the processes
\cite{Hein}: Pb+Pb $(T_f \approx 115~MeV)$ and S+S $(T_f \approx 145~MeV)$
at the SPS, in order to determine the critical temperature, $T_c \approx
145~MeV$,
and fix the parameter $I_o \approx 1.86$. Again, we notice that the
relatively
small value of the freeze-out temperature in Pb+Pb collisions (at the SPS)
is associated with the fact that in this process the system of quark matter
is not driven close to the critical point.
On the contrary the freeze-out temperature in S+S collisions remains
practically equal to the critical value, due to the proximity of this
process to the critical point, at the freeze-out stage (Fig.~1).
Finally, in Fig.~3, the rapidity distribution of net baryons in Au+Au
collisions at RHIC ($L \approx 11, z_c \approx 1.1$) is shown,
relying upon the fact that  in this process the distance from the
critical point is small ($d_c \approx 0.27$) and eq.(\ref{eq:eq11c}) is valid.

\section{Baryonic critical fluctuations}

Once the phase of quark matter has reached the critical point in
a particular class of experiments, as discussed in the previous section,
strong critical fluctuations are expected to form intermittency patterns both
in the pion and net-baryon sector. As already mentioned in the introduction,
the origin of these fluctuations can be traced in the presence, at $T=T_c$,
of a zero mass field with a classical profile ($\sigma$-field) which, under
the assumption of a phase transition in local thermal equilibrium, is
described by an effective action in $3-d$, the projection of which onto
rapidity space is written as follows \cite{Anto1}:

\begin{equation}
\Gamma_c \approx \frac{\pi R_{\perp}^2}{C_A} \int_{\delta y} dy \left[
\frac{1}{2} \left( \frac{\partial \sigma}{\partial y} \right)^2 + 2 C_A^2
\beta_c^4 \sigma^{\delta +1} \right]~~~;~~~C_A=\frac{\tau_c}{\beta_c},~~~
\beta_c=T_c^{-1}
\label{eq:eq12}
\end{equation}
Equation (\ref{eq:eq12}) gives the free energy of the $\sigma$-field within
a cluster of size $\delta y$ in rapidity and $R_{\perp}$ in transverse space.
The critical fluctuations generated by (\ref{eq:eq12}) in the pion sector
have been studied extensively in our previous work \cite{Anto1}, therefore,
in what follows, we are going to discuss the fluctuations induced by the
$\sigma$-field in the net-baryon sector, noting that a direct measurement
of these fluctuations may become feasible in current and future heavy-ion
experiments. For this purpose we introduce in eq.(\ref{eq:eq12}) the order
parameter $m(y)$ through the following equations:
\begin{eqnarray}
\sigma(y)\approx F \beta_c^2 m(y)~~~;~~~F \equiv -\frac{\lambda \langle
\bar{q}q \rangle_o}{\rho_c}~~~;~~~\langle \bar{q} q \rangle_o \approx
-3~fm^{-3} \nonumber \\
\Gamma_c \approx g_1 \int_{\delta y} dy \left[ \frac{1}{2} \left(
\frac{\partial \hat{m}}{\partial y} \right)^2 + g_2 \vert \hat{m}
\vert^{\delta +1} \right]~~~~~;~~~~~\hat{m}(y)=\beta_c^3 m(y) 
\label{eq:eq13}
\end{eqnarray}
where: $g_1=F^2\left(\frac{\pi R_{\perp}}{C_A \beta_c^2}\right)$,
$g_2=2 C_A^2 F^4$. The partition function $Z=\int {\cal{D}}[\hat{m}]
e^{-\Gamma_c[\hat{m}]}$ for each cluster is saturated by instanton-like
configurations \cite{Anto2} which for $\delta y \leq \delta_c$ lead to
self-similar structures, characterized by a pair-correlation function of the
form:
\begin{equation}
\langle \hat{m}(y) \hat{m}(0) \rangle \approx \frac{5}{6}
\frac{\Gamma(1/3)}{\Gamma(1/6)} \left(\frac{\pi R_{\perp}^2 C_A}{\beta_c^2}
\right) F^{-1} y ^{-\frac{1}{\delta +1}}
\label{eq:eq14}
\end{equation}
The size, in rapidity, of these fractal clusters is $\delta_c \approx
\left(\frac{\pi R_{\perp}^2}{16 \beta_c^2 C_A^2}\right)^{2/3}$ according to
the geometrical description of the critical systems \cite{Anto2}. Integrating
eq.(\ref{eq:eq14}) we find the fluctuation $\langle \delta n_b \rangle$ of
the net-baryon multiplicity with respect to the critical occupation number
within each cluster, as follows:
\begin{equation}
\langle \delta n_b \rangle \approx F^{-1}
\left(\frac{\pi R_{\perp}^2 C_A}{2 \beta_c^2}\right)
\frac{2^{2/3} \Gamma(1/3)}{\Gamma(1/6)} \delta_c^{5/6}
\label{eq:eq15}
\end{equation}
The dimensionless parameter $F$ is of the order $10^2$ and the size
$\delta_c$, on general grounds $(R_{\perp} \stackrel{<}{\sim} 2 \tau_c)$ is
of the order of one $(\delta_c \stackrel{<}{\sim} 1)$.
This is in agreement with the fact that the rapidity separation of two
causally correlated space-time events is logarithmicaly bounded,
$\delta y \leq \ln\left(1+\frac{\delta \tau_c}{\tau_c}\right)$, leading,
practically, to a direct
correlation length of the order of one ($\delta y \stackrel{<}{\sim} 1$).
However, in reality, the global baryonic system  develops fluctuations at all
scales in rapidity since the direct correlation (O($\delta_c$)) propagates
along the entire system through the cooperation of many self-similar clusters of
relatively small size ($\delta_c \approx 0.35$ and $\langle \delta n_b
\rangle \approx 140$). We have quantified this mechanism in a Monte-Carlo
simulation for the conditions of the experiments at RHIC (in RHIC the size
of the system is $L \approx 11$) in
order to generate baryons with critical fluctuations. The distribution of
the order parameter $\vert m(y) \vert$ describing the fluctuations of the
``critical" baryons in the rapidity space for a typical event as well
as the corresponding intermittency analysis in terms of factorial moments are
presented in Fig.~4. The intermittency exponent of the second moment $F_2$ in
rapidity is found to be $s_2 \approx 0.18$ which is very
close to the theoretically expected value ($\frac{1}{6}$) of a monofractal
$1-d$ set with fractal dimension $\frac{5}{6}$. The
quantitative details of this new class of phenomena, concerning critical
fluctuations in the net-baryon sector, and in particular the intermittency
pattern of the observable net-baryon density, go beyond the purpose of this
Letter and will be presented elsewhere \cite{Anto3}.

\section{Concluding remarks}

In conclusion, we have shown that the baryonic sector in an experiment with
heavy ions, possess valuable information regarding the proximity and
observation of the second order QCD critical point.
Complementary suggestions for the significance of net-baryon fluctuations
in order to trace the critical line of first order in the phase diagram are
described in \cite{Gavin}. In particular the
measurements of net-baryon spectra in
rapidity provide a valuable set of obsevables in heavy ion experiments,
in connection with the phenomenology of the QCD critical point. The trend
of the existing data at the SPS suggests the presence of a critical point
of second order in the phase diagram (Figs.~1,2) specified by the critical values
of temperature and density: $T_c \approx 145~MeV$, $\rho_c \leq \frac{\rho_o}
{5}$. A scaling law for the net-baryon density at midrapidity $n_b$,
as a function of the initial energy and the number of participants, has
been established in the neighbourhood of the critical point. The scaling
function incorporates the indices of the universality class (critical
exponents) whereas the scaling variable gives a measure of the proximity
of a given experiment to the critical point. On the basis of this
investigation, the experiments at RHIC are very likely to reach the QCD
critical point and as a first sign of this new phenomenon we have predicted
a rather unconventional profile of net-baryon density in rapidity
(in Au+Au collisions), associated with the presence of a critical point,
nearby (Fig.~3). Finally we have indicated the presence of strong
intermittency effects in the rapidity spectrum of the net baryons which can
in principle be used to reveal experimentally the QCD critical point and
confirm its universality class (Fig.~4).


{}
\vspace*{1cm}
\begin{center}
{\bf{Figure Captions}}
\end{center}
\vspace{0.5cm}

\noindent
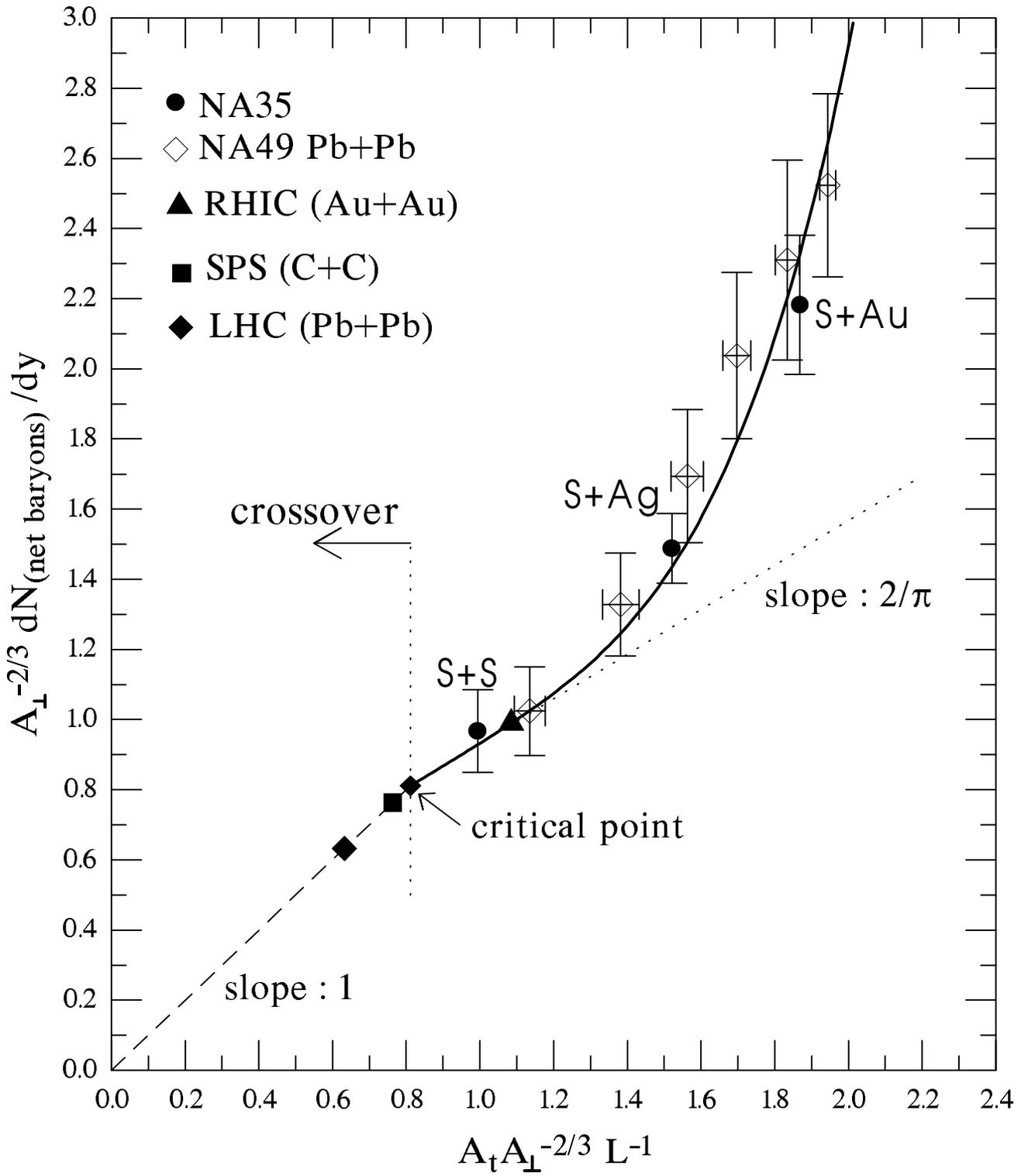
FIG.~1. The scaling law (10) is illustrated together
with measurements at the SPS. The critical point and the corresponding
break in the slope of $\Psi(z_c,\rho_c)$ are also shown.

\noindent
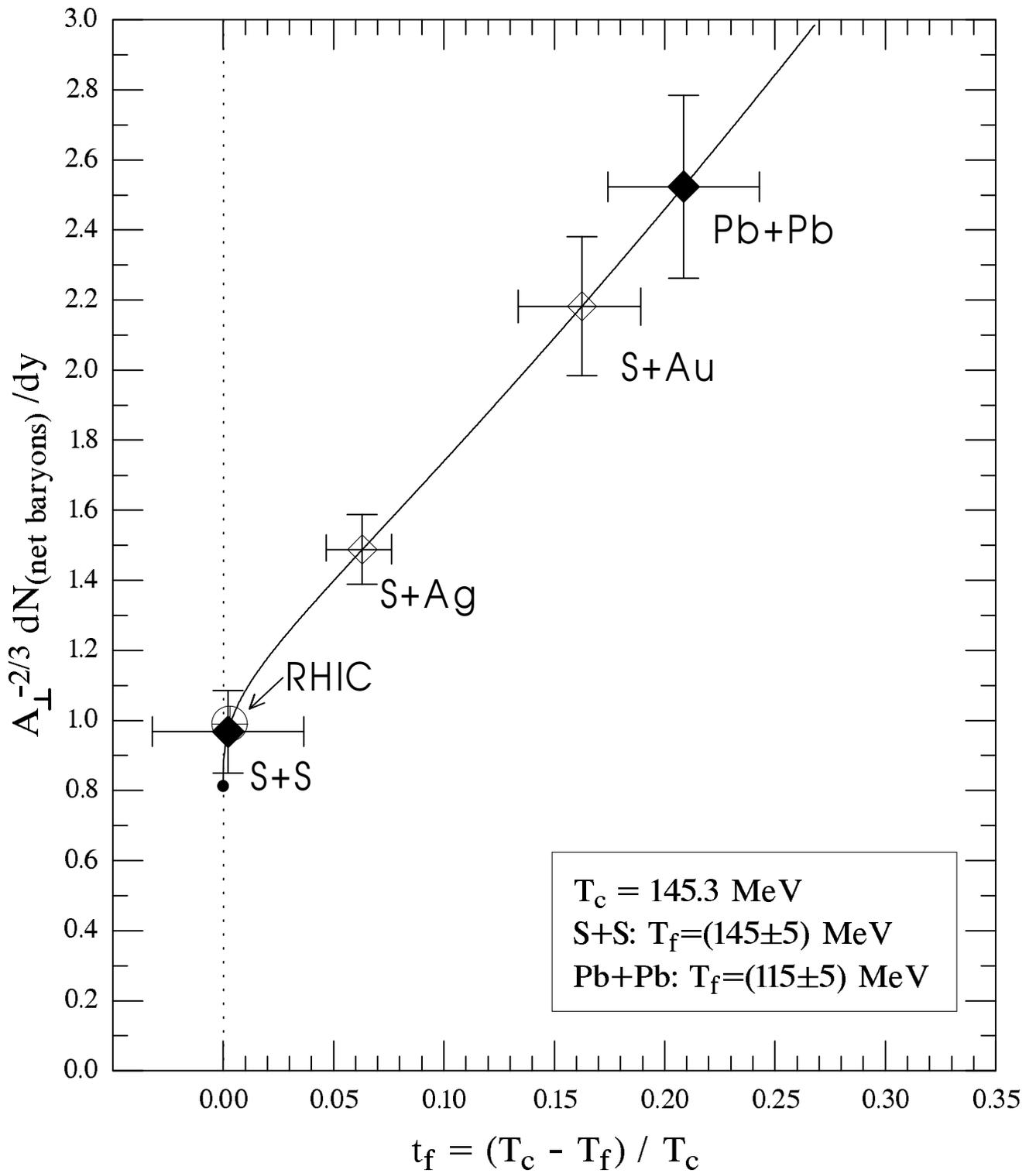
FIG.~2. The freeze-out line, $n_b$ as a function of $t_f$, is
illustrated together with measurements at the SPS (S+S, Pb+Pb). The
extracted freeze-out points for S+Au and S+Ag are also shown in this diagram.

\noindent
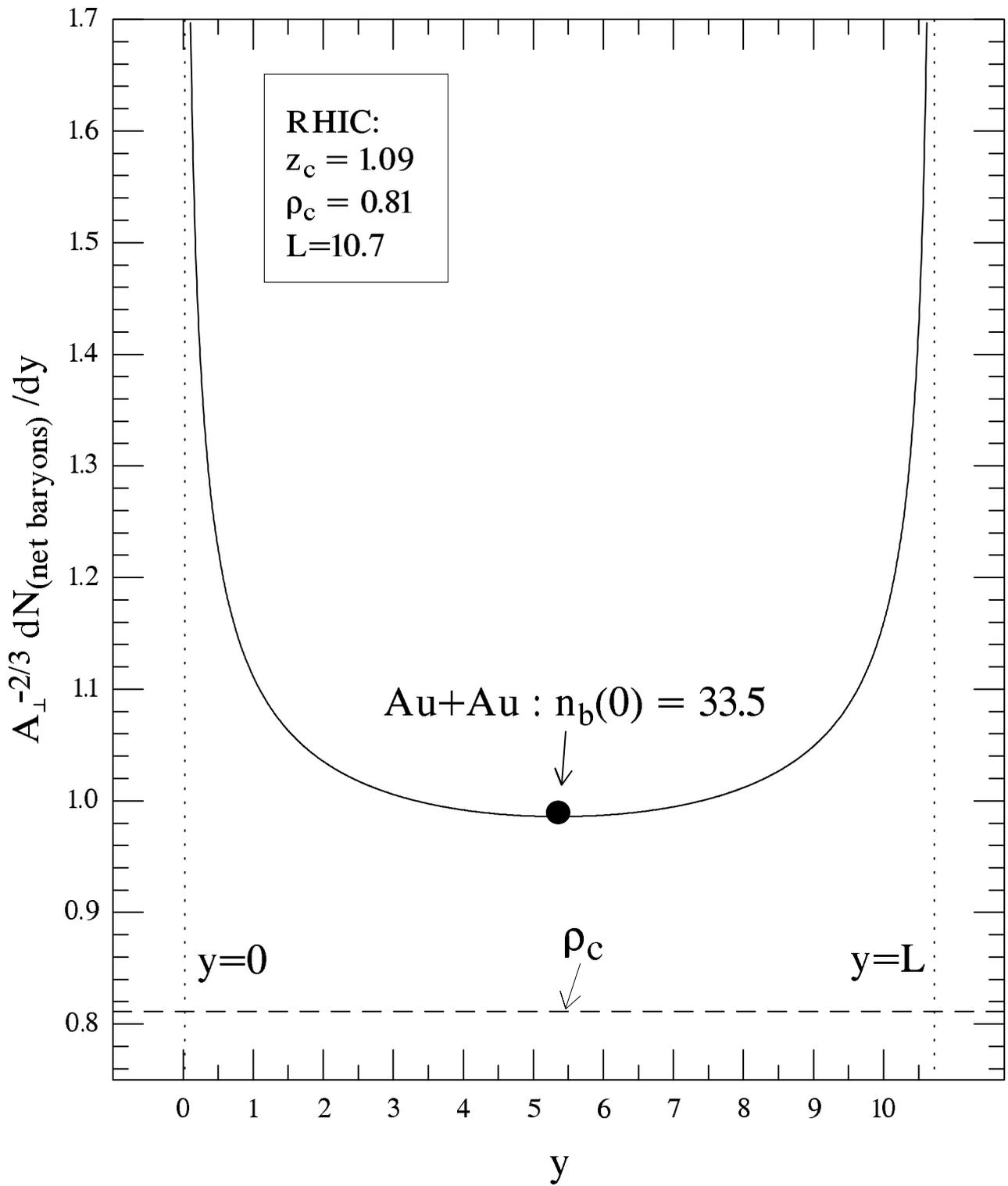
FIG.~3. The net-baryon profile in rapidity for Au+Au collisions at
RHIC energies is shown, as predicted by eq.(\ref{eq:eq11c}).

\noindent
FIG.~4. (a) The distribution of $\vert m(y) \vert$ in rapidity for a
MC-generated event. (b) The first three factorial moments for the event shown
in (a) in a log-log plot. A linear fit determining the slope $s_2$
($\approx 0.18$) of the second moment is also shown.

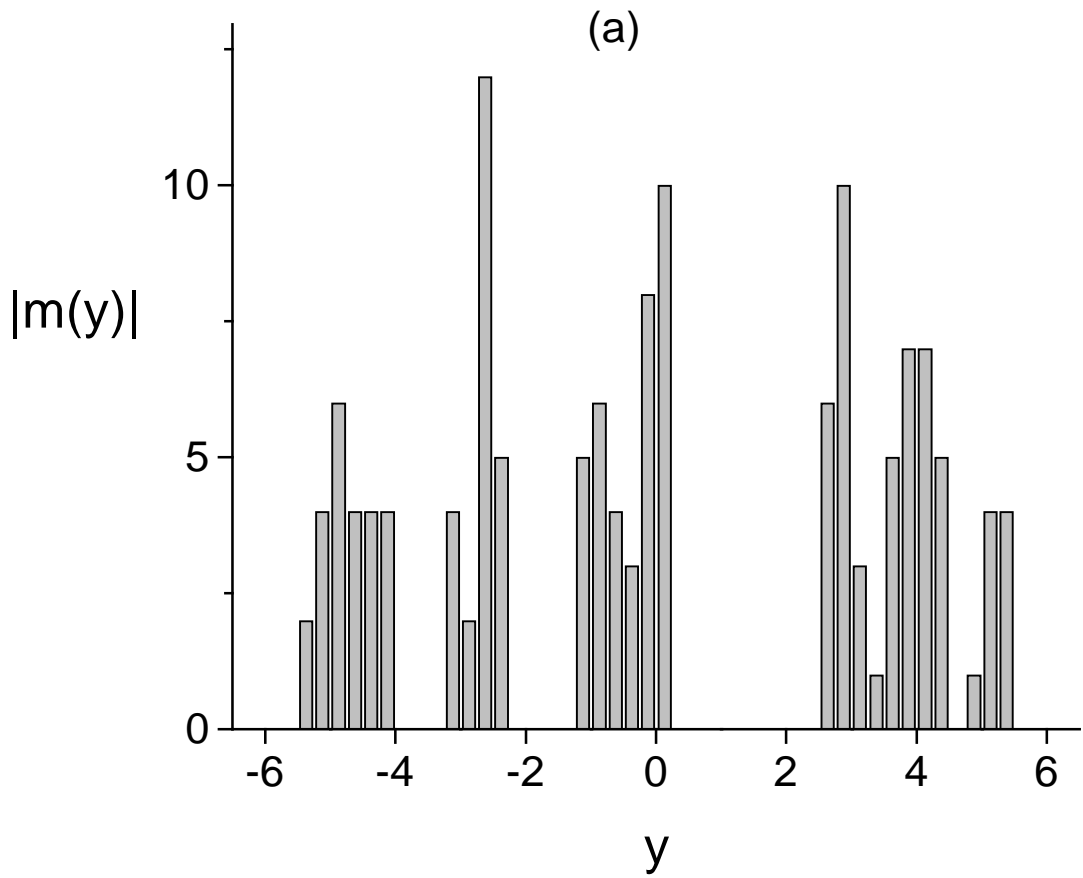
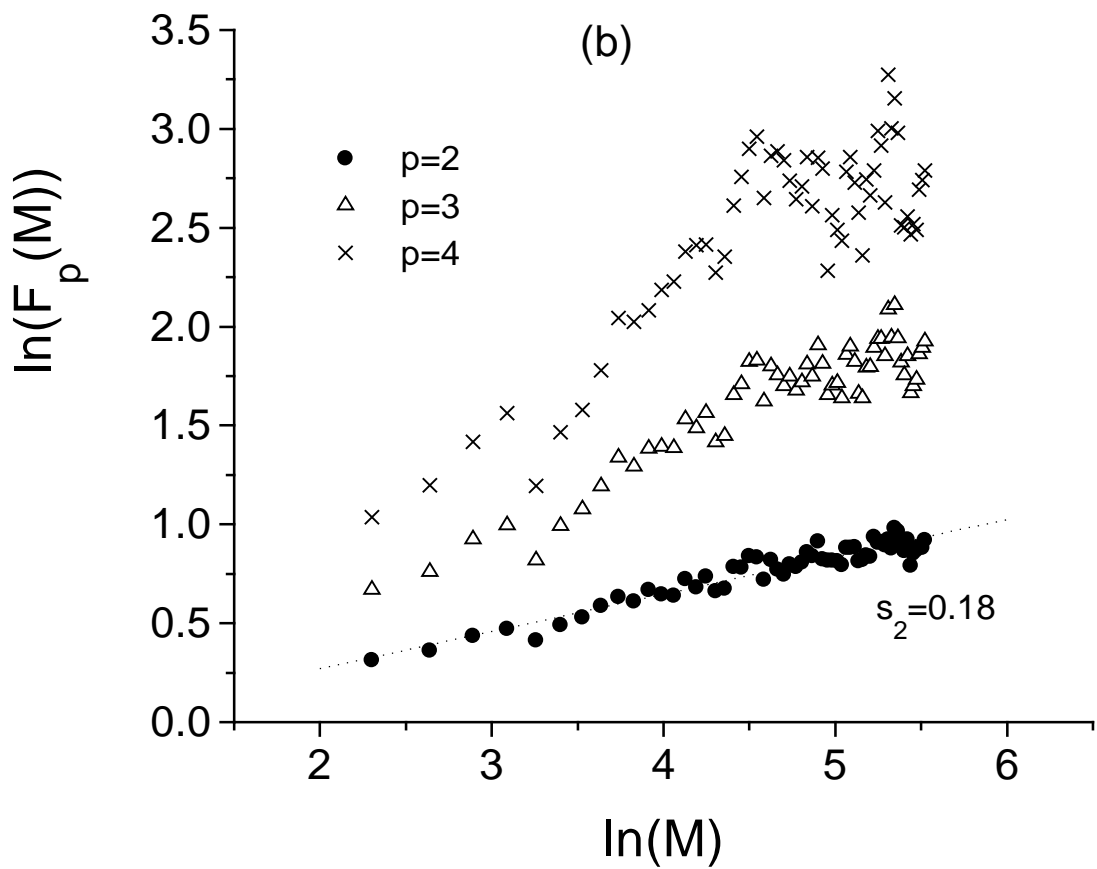
\end{document}